\documentclass[conf]{new-aiaa}
\usepackage[utf8]{inputenc}
\usepackage{graphicx,epstopdf}
\usepackage{amsmath}
\usepackage[version=4]{mhchem}
\usepackage{siunitx}
\usepackage{longtable,tabularx}
\usepackage{xcolor}
\usepackage{gensymb}
\usepackage{subfig}
\usepackage{fancyhdr}
\title{High fidelity simulations of unstart phenomena in a scramjet inlet due to angle of attack}
\author[1]{Jeremy Redding}
\author[2]{Emma Cavanaugh} 
\author[3]{Luis Bravo}
\author[3]{Muthuvel Murugan}
\author[2]{Venkat Narayanaswamy} 

\affil[1]{Aerospace Engineering Department, University of Cincinnati, Cincinnati, OH, USA}
\affil[2]{Mechanical \& Aerospace Engineering Department, North Carolina State University, Raleigh, NC, US}
\affil[3]{DEVCOM Army Research Laboratory, Aberdeen Proving Ground, MD, USA}

\begin{document}
\maketitle
\begin{abstract}
     This work investigates the unsteady behavior of unstart phenomena within a scramjet inlet using advanced computational techniques. Scramjets and ramjets, with their reliance on inlet compression, offer promising airbreathing propulsion for hypersonic regimes. This research focuses on understanding and modeling the onset of unstart phenomena in supersonic inlets, a critical step towards developing mitigation strategies. These strategies have the potential to improve engine efficiency, range, and maneuverability of hypersonic vehicles. To achieve this, the state-of-the-art compressible flow solver, Eilmer, is used to simulate shockwave behavior within the inlet/isolator of a planar scramjet characterized experimentally at North Carolina State University (NCSU). The simulations employ the classical Spalart Almaras, Reynolds Averaged Navier-Stokes (RANS), turbulence model to capture the dynamics of shockwaves during unstart. Baseline comparisons are presented with the wind tunnel experiments via the shock structures present within the isolator section conducted at Mach 3.9 on a 3D scramjet inlet model. Simulations are then carried out at varying angles of attack (0 to 10 deg) and multiple pitch rates (10 deg/sec and 100 deg/sec) to demonstrate the shock train inertial response and to characterize unstart onset comprehensively. In both cases the timing of inlet unstart is observed to correlate well with the rapid surge in exit pressure as well as shock detachment at the lower leading edge region. Lastly, exit pressures are significantly higher in the 10 deg/s case than in that of the 100 deg/s case at the same angle of attack. These observations suggest that unstart is not only dependent on angle of attack but also on AoA pitch rate. The findings provide valuable insights into the unsteady flow behavior during hypersonic inlet unstart, with potential applications for unstart detection at high angles of attack.
\end{abstract}

\section{Introduction}

Scramjets offer the potential for revolutionary hypersonic air-breathing propulsion due to their ability to operate efficiently for a long range, time critical capability. However, a critical challenge to their stable operation lies in a phenomenon known as unstart. Unstart disrupts the supersonic airflow entering the scramjet inlet, leading to a drastic reduction in air mass flow for combustion and a subsequent loss of thrust and vehicle control. This instability not only limits engine performance but also introduces significant unsteady mechanical loads on the scramjet structure. Researchers have extensively investigated the physics behind unstart and developed various control strategies \cite{burke_aiaa_2021}\cite{devaraj_ast_2021}\cite{im_pas_2018}. Boundary layer bleeding, a technique where a portion of the slow-moving air near the inlet walls is removed, has been shown to be effective in preventing boundary layer separation and subsequent unstart \cite{chang_AJ_2009}\cite{kang_AIAAJ_2020}. Additionally, studies have explored the use of variable geometry inlets that can dynamically adjust the inlet flow path to optimize shock wave interaction and prevent unstart under varying flight conditions\cite{reardon_AIAAJ_2021}\cite{starkey_aiaa_2005}. Computational simulations are also playing a crucial role in understanding the complex flow dynamics within the inlet and predicting unstart boundaries. Further research is ongoing to develop robust and adaptable control mechanisms for unstart prevention, paving the way for a new generation of reliable hypersonic vehicles powered by scramjet engines.
\\

\begin{figure}[hbt!]
    \centering
    \includegraphics[width=0.75 \textwidth]{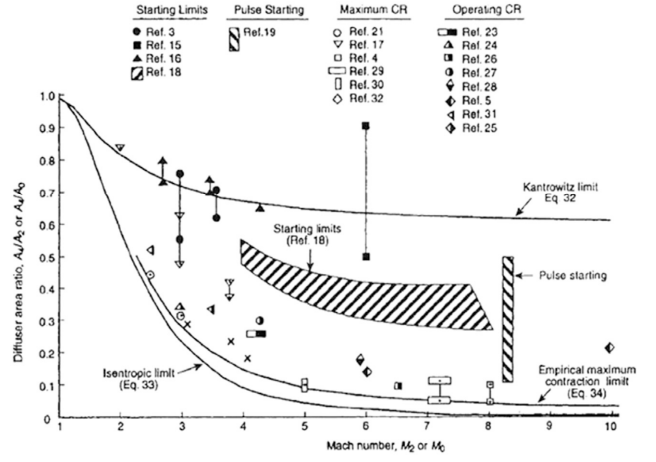}
    \hspace{10mm}
    \caption{Inlet starting regimes for various Ma and contraction ratios. Kantrowitz, isentropic, and empirical maximum contraction ratio limits for inlet starting are shown \cite{im_pas_2018}}
    \label{fig:Kantrowitz}
\end{figure}

Understanding the key factors influencing inlet unstart is crucial for ensuring stable and efficient operation. Extensive research has been conducted to delineate the boundaries for unstart based on freestream conditions and vehicle geometry. Among these parameters, freestream Mach number (Ma) and the inlet Compression Ratio (CR) have received the most attention for defining unstart boundaries. For instance, the Kantrowitz limit establishes a critical inlet CR that can trigger unstart based on the freestream Mach number and the ratio of specific heats of the incoming airflow \cite{chang_AJ_2009}. Essentially, if the inlet CR of a scramjet exceeds this critical value, the excessive compression occurring within the inlet can lead to unstart. However, relying solely on freestream Mach number, as in the Kantrowitz limit, to define a universal critical CR for unstart at a given Mach number proves impractical. This is because real-world scenarios introduce complexities, such as three-dimensional (3D) non-axisymmetric inlets, where CR alone is insufficient for defining the starting/unstarting criterion \cite{im_pas_2018}\cite{anderson_aiaa_1991}\cite{devaraj_ast_2021}. This limitation arises because the critical condition depends not just on the CR but also on the intricate interplay between the inlet and isolator geometries (as depicted in Figure~\ref{fig:Kantrowitz} ). As evident in the figure, the Kantrowitz limit falls short as a standalone criterion. The 3D geometry of the inlet along the flow path dictates the compression process, which significantly impacts the resulting stagnation pressure loss as the flow contracts through the inlet. Another important real-world complexity is the angle of attack the vehicle experiences during its hypersonic flight trajectory. While design conditions for steady cruise flight aims for close to zero angle of attack, excursions can occur due to sudden maneuvers. Understanding how varying angles of attack influence unstart characteristics is essential for robust engine design and overall flight stability.\\

Although a significant amount of work has been aimed at understanding unstart, very little is known about the parameters governing the unstart or restart dynamics due to variations in vehicle angle of attack. To address this shortcoming, the objective of the present study is to computationally investigate the flow physics of a scramjet inlet for cases of normal running and unstart conditions. Inlet flow physics is experimentally compared at Mach 3.9 based on the North Carolina State University (NCSU) wind tunnel facility that is equipped with a 3D scramjet inlet/isolator model with a shock on lip design condition at Mach 2.5 and a physical area contraction ratio, CR, 2.64. This is followed by simulations at varying angles of attack (AoA) to better characterize the onset of unstart dynamics. The results provide valuable insights into the parameters governing unstart shocktrain dynamics to enhance flight performance. The remainder of the paper is organized as follows, Section 1 presented the motivation and overall objective of this work, Section 2 presents the computational framework including a description of the numerical and modeling capabilities available in Eilmer hypersonic solver, Section 3 presents the results and discusses the influence of angle of attack at different pitch rates on unstart, and finally Section 4 presents the summary and future work.


\section{Methods}

\subsection{Eilmer computational fluid dynamics solver}

Our investigation utilizes Eilmer, a general purpose open-source CFD solver developed by the University of Queensland, to model the intricacies of high Mach number internal flow physics. It is written in the D programming language and uses an embedded Lua interpreter for configuration and run-time customization. Eilmer contains a wide range of customizable options, including high temperature gas models, turbulence models, and flux calculators. Multi-physics simulation capability includes fluid-structure interactions, fluid-thermal interactions, and fluid-radiation coupling \cite{gibbons_cpc_2023}. Eilmer excels at handling compressible flows within 2D and 3D geometries thanks to its multi-block structured grid approach as well as its ability to handle unstructured grids. Eilmer leverages explicit time-stepping to achieve time-accurate solutions, making it ideal for capturing unsteady phenomena. It offers a comprehensive suite of turbulence models (including Spalart-Allmaras, Baldwin-Lomax, k-omega, etc) and thermochemical models, even incorporating finite-rate chemistry for simulating non-equilibrium flows as well as thermal energy exchange models. The solver has been deployed on the DoD high performance computing (HPC) platforms including Nautilus, Narwhal and Warhawk with acceptable scalability over 1000s of CPU cores. Further, it's accuracy has been extensively validated through various hypersonic test cases within our group including, \cite{redding_pof_2023}\cite{gamertsfelder_asme_2023}.\\

In the current simulations, unsteady 2D RANS with the Spalart Allmaras model is invoked. The initial conditions are: 

\begin{table}[!h]
\centering
\begin{tabular}{|c|c|c|}
	\hline
	Variable & Value  & Units \\
	\hline\hline
	Temperature & 74.22 & (K)  \\
	\hline
	Pressure & 9,556.0 & (Pa) \\
	\hline
	$U_{\infty}$ & 673.488 & (m/s)\\
	\hline

\end{tabular}
\caption{Initial conditions in the simulation}
\label{tab:conds}
\end{table}

\subsection{NCSU wind tunnel facility}
The scramjet inlet experimental article was investigated in a variable Mach number supersonic blowdown wind tunnel facility at North Carolina State University. This supersonic wind tunnel operates between Mach 1.4 and 4; for the baseline work, the freestream Mach number was set at 3.9, which is well above the inlet design Mach number of 2.5. Experiments are run for 8 seconds with the middle six seconds exhibiting stable freestream pressure conditions. \\

\begin{figure}[hbt!]
    \centering
    \includegraphics[width=0.5 \textwidth]{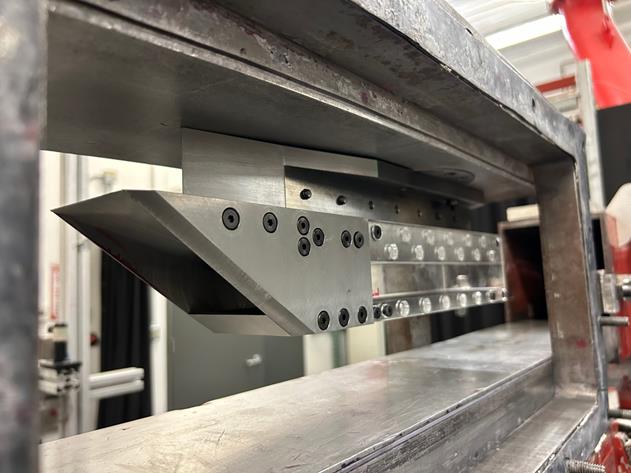}
    \hspace{10mm}
    \caption{Inlet model, attached isolator, and acrylic wall configuration of the NCSU model}
    \label{fig:isolatorscram}
\end{figure}

Experiments are analyzed via the shock structures present within the isolator section. Schlieren imaging techniques are employed for this visualization. Imaging of the internal shock structure of the isolator is made possible through the use of two clear acrylic side walls attached to the supersonic wind tunnel and isolator section of the test article. These walls are oriented perpendicular to the flow in order to allow for profile view imaging of the shock structures within the isolator section.  An image of the test article is shown in Figure ~\ref{fig:isolatorscram}. The wind tunnel acrylic walls were removed in this figure. For the Schlieren imaging set up, a white LED light source was used and oriented in a z-shape configuration with the mirror reflecting the light beam, acrylic side walls of the supersonic wind tunnel, acrylic walls of the isolator, and mirror reflecting the light into the high-speed camera oriented in parallel. The images presented in this paper were recorded at 10000 fps with a shutter speed of 1/30000 seconds.\\

\subsection{Scramjet inlet design}
The test article comprises a 3D inlet/isolator model that has a shock-on-lip (SOL) design condition at Mach 2.5. The model was made of steel and had a rectangular cross-section that had an inlet capture height of 2.64 inches, a width of 3 inches, and a total length of 16.5 inches. The isolator formed 10 inches of the total length and the inlet had a physical area contraction ratio, CR, of 2.64. For these tests shown in the following results the inlet and isolator were operated at the above design condition of Mach 4.0. In light of this, the SOL condition is not met and instead the bow shock rests slightly inside the cowl. \\


\section{Results and Discussions}

\subsection{Baseline analysis and comparison with experiments}

\begin{figure}[!h]
  \centering
  \includegraphics[width=0.75\linewidth]{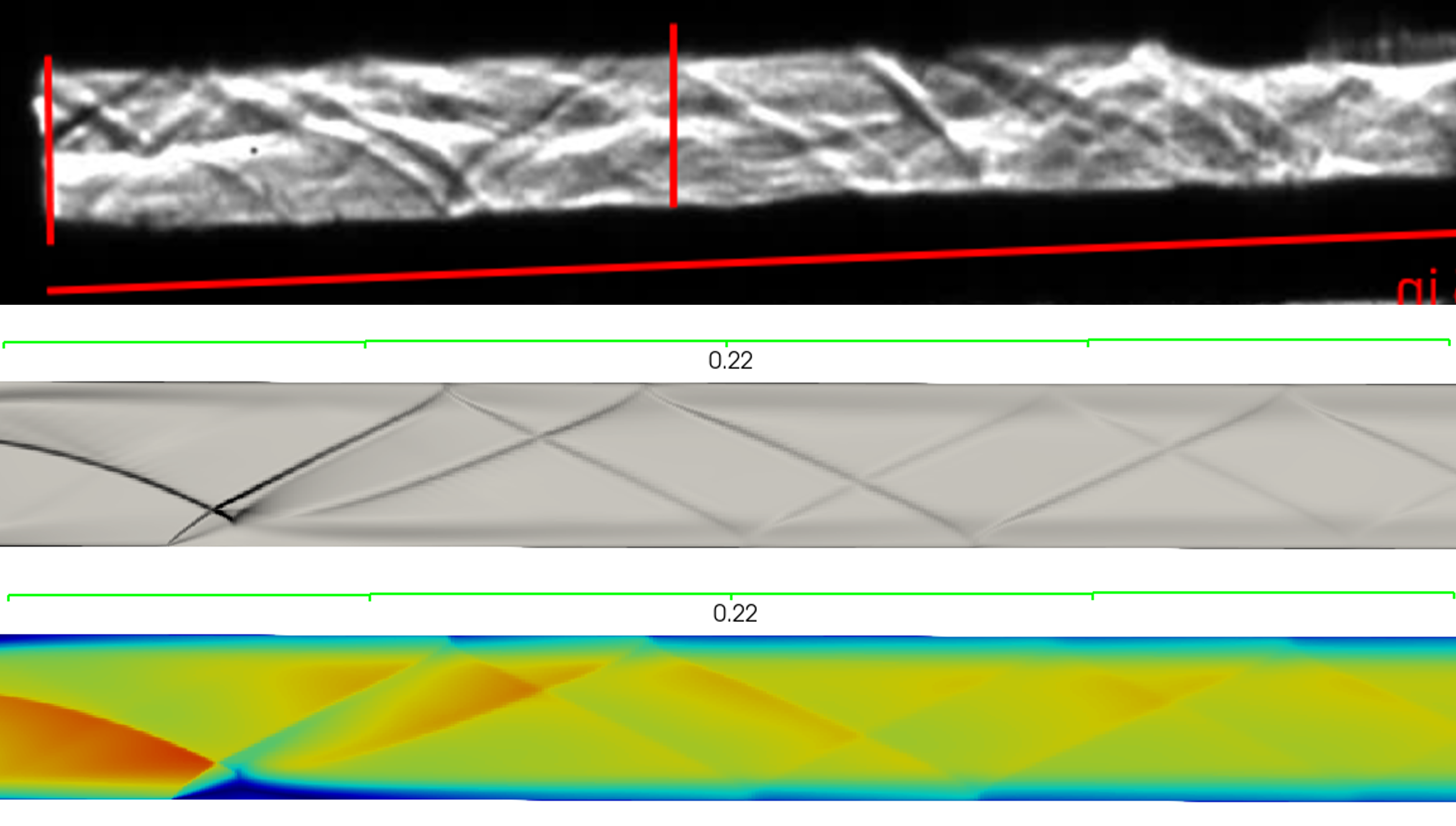}
  \caption{Schlieren of the experimental run in a fully started state (top), numerical schlieren (middle) and mach number contours (bottom) at zero angle of attack.}
  \label{fig:gradients}
\end{figure}

Basic qualitative comparison is made between the numerical and experimental works in \autoref{fig:gradients}. While the basic shock structure is captured as being an x-type shock structure, the experimental and numerical methods may need to be further refined for better detailed alignment between the two. The application of other RANS turbulence closures can change the shock train dynamics and dissipative characteristics and this will be investigated in our future work. Here, grid sizing is based on achieving a y+ = 1.0 for the current conditions.  \\

\subsection{Angle of attack rates and range}

Inlet unstart caused by angle of attack change is dependent on the rate of change of angle of attack. In a sudden maneuver, inertial motion and subsequent response of the shock train may drastically affect attainable angles of attack, or even potentially keep the shock train within the isolator (if it cannot respond to flow dynamics quickly enough). \\

In order to investigate these limits, and begin to generalize the flow physics contained within this phenomena, two 2D unsteady simulations are performed at two angle of attack rates $\alpha(t)$. The two cases include:

\begin{equation} \label{eq1}
\begin{cases}

   \text{Case 1,} & \alpha_1(t) = 10 \degree \text{/s} \\
   \text{Case 2,} & \alpha_2(t) = 100 \degree \text{/s}

    \end{cases}
\end{equation}

In each of these cases, the values range from $\alpha = 0\degree$ to $10\degree$. The reason for selecting this angle of attack range is based on prior experience with this inlet, noting unstart at around ~8-10 degrees for statically changing tests. \\

\subsection{Measurement locations and shocktrain physical characteristics}

\begin{figure}[!h]
  \centering
  \includegraphics[width=1\linewidth]{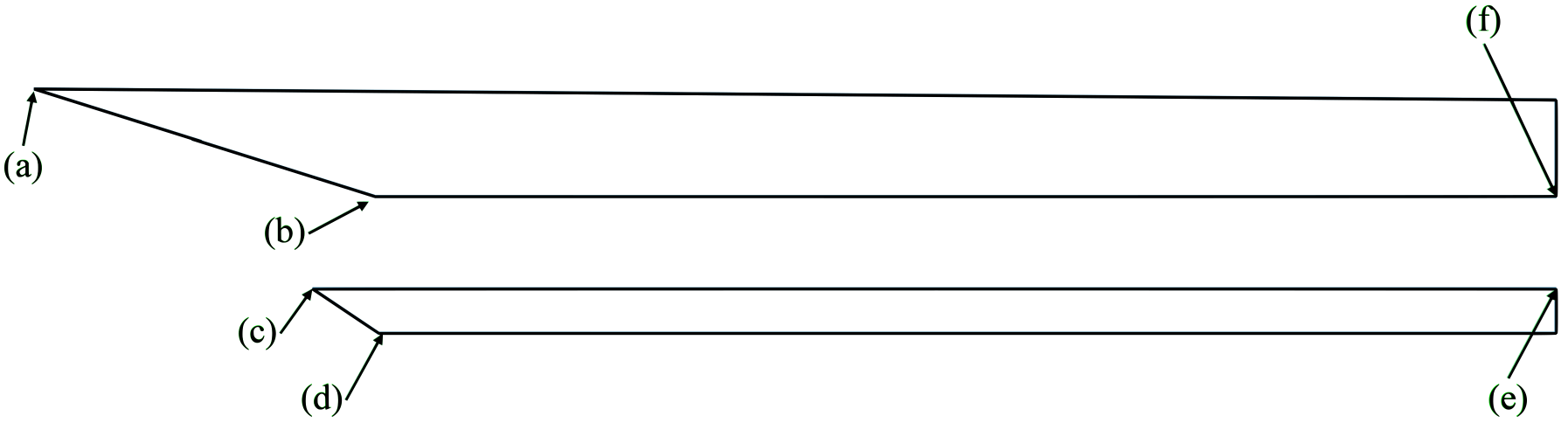}
  \caption{Geometry edge locations, the coordinates and naming convention of which is given in \autoref{tab:locations}}
  \label{fig:LocationsFig}
\end{figure}

 In the subsequent discussions, the points shown in \autoref{fig:LocationsFig} will be referred to by the names given in \autoref{tab:locations}\\

\begin{table}[!h]
\centering
\begin{tabular}{|c|c|c|c|}
	\hline
	Points & Name  & $x$ (m) & $y (m)$ \\
	\hline\hline
	(a) & Upper Leading Edge (ULE) & 0.05715 & 0.117056 \\
	\hline
	(b) & Upper Inlet Entrance (UIE) & 0.1524 & 0.087846\\
	\hline
	(c) & Lower Leading Edge (LLE) & 0.13335 & 0.0627\\
	\hline
	(d) & Lower Leading Bypass (LLB) & 0.1524 & 0.05\\
	\hline
	(e) & Lower Exit (LE) & 0.47625 & 0.0627\\
	\hline
	(f) & Upper Exit (UE) & 0.47625 & 0.087846\\
    \hline 
\end{tabular}
\caption{Coordinates and names for labeled locations given in \autoref{fig:LocationsFig}}
\label{tab:locations}
\end{table}

As the supersonic flow enters the inlet, a x-type shock train is formed by the angle of the upper surface. The shock initiated at ULE enters into the isolator, impinging with the lower wall of the isolator and reflecting. This is the source of one of the two shock trains. The other shock train is initiated from the shock attached to point LLE. The interaction between these two shock structures and their subsequent reflections down the isolator generates an unsteady x-type shocktrain. Expansion of the flow as it crosses point UIE causes a recirculation region at the beginning of the isolator. This recirculation region appears to modify the shape and offset of the two interacting shockwaves, causing further unsteady behavior in the x-type shock train structure, at certain times causing the two to overlap appearing as a standard shock train. \\

\begin{figure}[htb]
  \centering
  \subfloat[Pressure ratio over time at isolator exit]{\includegraphics[width=0.5\textwidth]{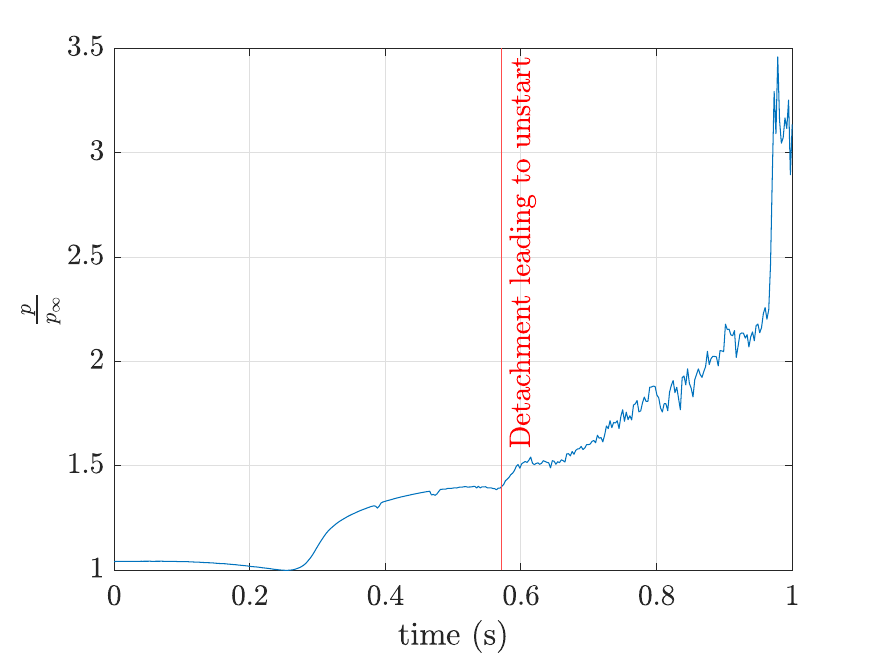}\label{fig:pvst10s}}
  \hfill
  \subfloat[Angle of attack verses oblique angle measured as in \autoref{fig:a10_61}]
  {\includegraphics[width=0.5\textwidth]{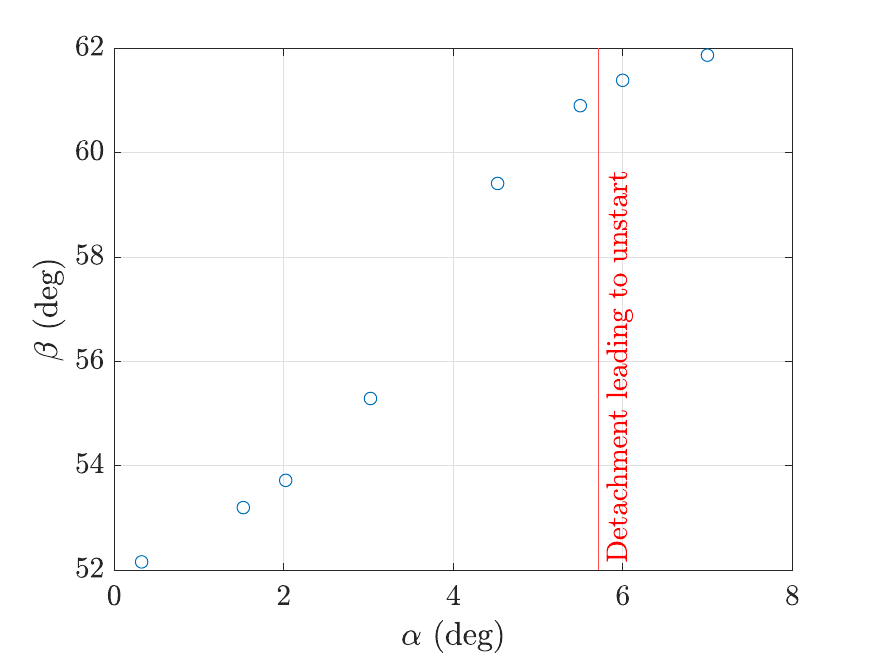}\label{fig:ab10s}}
  \caption{Predicted pressure and angle of attack measurements for $\alpha_1(t)$ = 10 deg/s}
\end{figure}
\begin{figure}[htb]
  \centering
  \subfloat[Pressure ratio over time at isolator exit]{\includegraphics[width=0.5\textwidth]{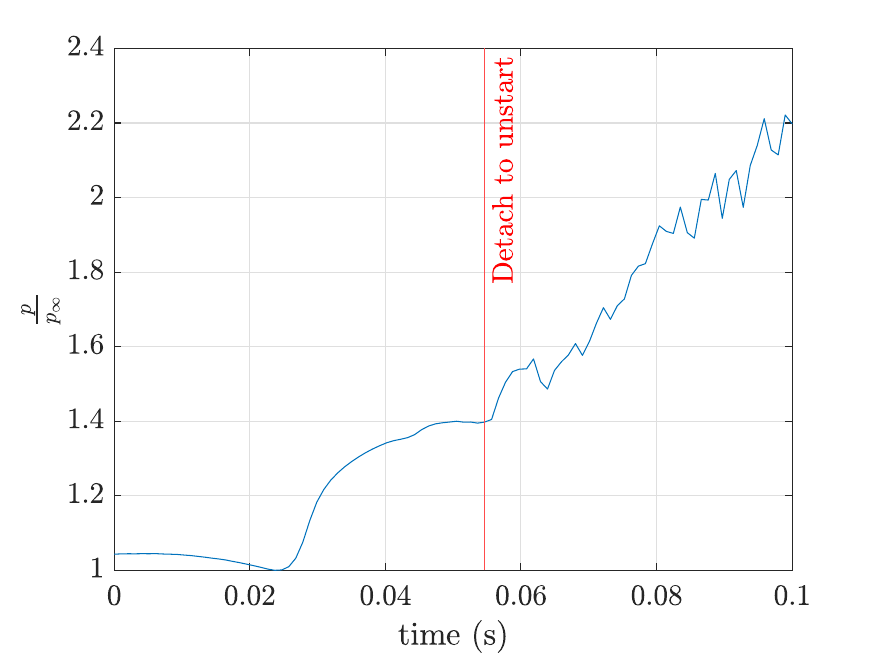}\label{fig:pvst100s}}
  \hfill
  \subfloat[Angle of attack verses oblique angle measured as in \autoref{fig:a10_61}]{\includegraphics[width=0.5\textwidth]{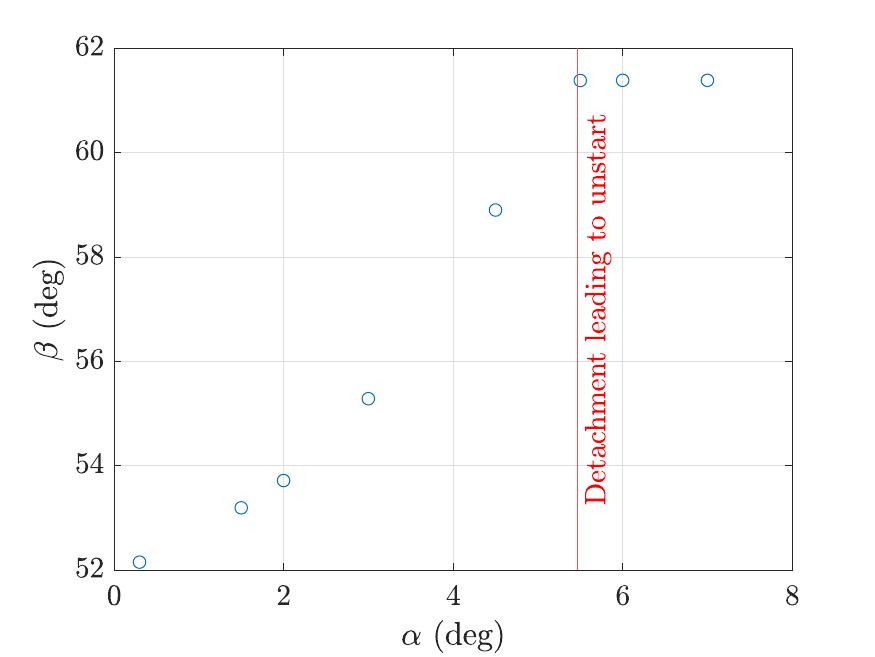}\label{fig:ab100s}}
  \caption{Predicted pressure and angle of attack measurements for $\alpha_1(t)$ = 100 deg/s}
\end{figure}

In the two tested cases, inlet unstart is observed close to the same time as the attached shock at point LLE becomes detached. In figures \autoref{fig:pvst10s}, \autoref{fig:pvst100s}, the pressure ratio between pressure near the exit of the isolator and the freestream pressure is plotted over time. For the 10 degree per second case, detachment and beginning of unstart occurs at around t = 0.57 seconds (57\% of the total time), and an angle of attack of $5.7\degree$. In the 100 degree angle of attack case, however, this same phenomena begins around t = 0.0546 seconds (54.6\% of the total time), and an angle of attack of $5.46\degree$. In this case, the high rate of change causes unstart at an earlier time, showing a dependence on shock train inertia. It is also noted that even though both simulations run to a maximum of $\alpha = 10\degree$ by their respective ending time, the pressure rise due to the beginning of unstart is not nearly as high. This indicates that not only is unstart dependent on the inertial motion of the leading shock in the shock train, but also the compressive, spring-like motion in the shocktrain itself. \\

In figures \ref{fig:ab10s}, \ref{fig:ab100s}, the time step for the corresponding unstart event as observed in figures \autoref{fig:pvst10s}, \autoref{fig:pvst100s}, is drawn with the red line. Shock detachment from the lower leading edge occurs early in the onset of unstart. In the 100 degree/s case, the oblique shock angle becomes a constant immediately at unstart. Only a few time steps are measured after this point as the shock is detached, affecting the accuracy of measurement.

\clearpage
\subsection{Visualizations of unstart process due to angle of attack}

\begin{figure}[!h] 
  \centering
  \includegraphics[width=\linewidth]{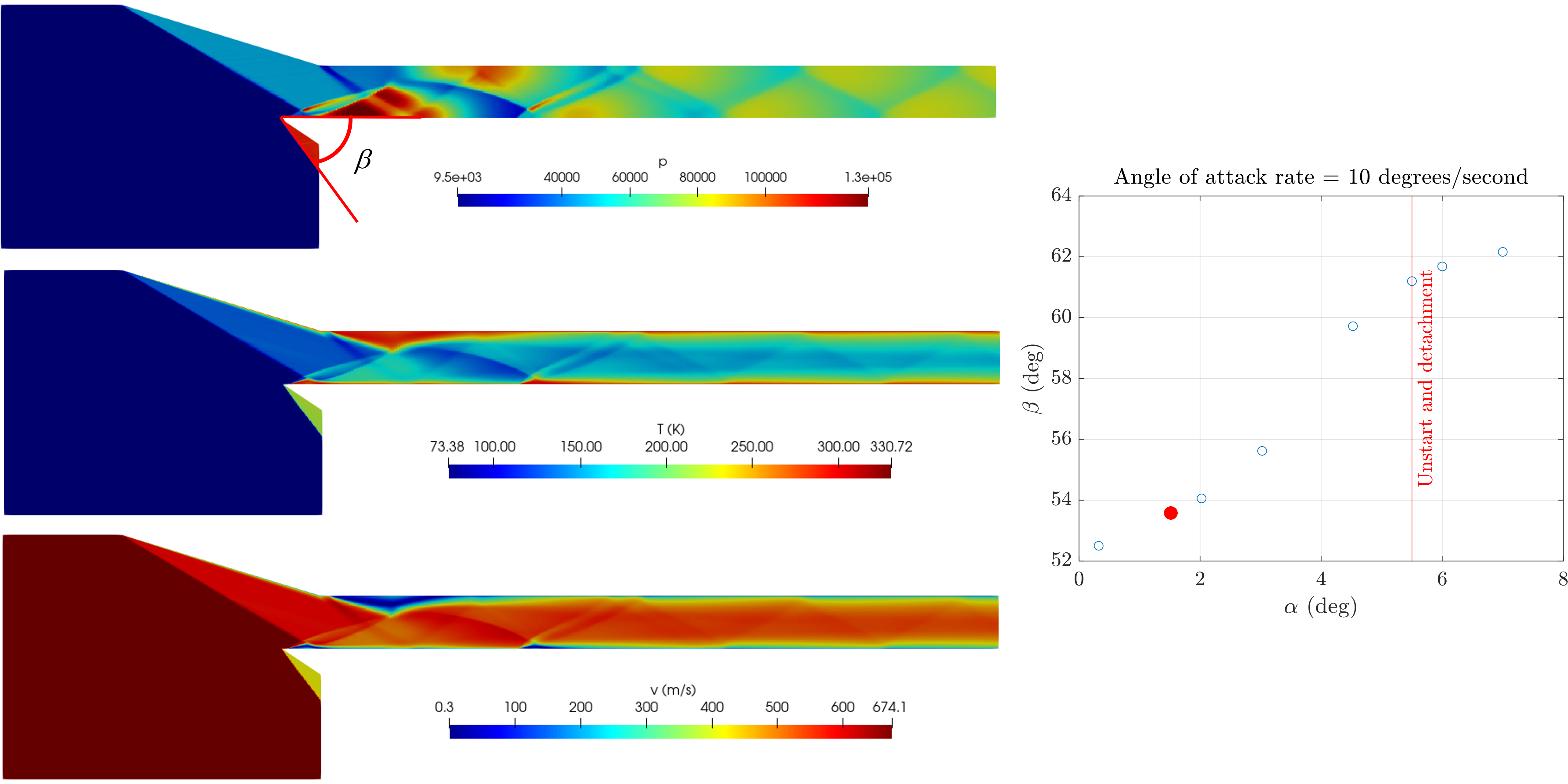}
  \caption{For the 10 deg/second angle of attack case, the pressure (top), temperature (middle), and velocity (bottom) contours at $\alpha = 1.5 \degree$ (red dot on the plot); relative oblique angle $\beta$ measured at the location in the figure, the shock generated from the point LLE, given on the right}
  \label{fig:a10_61}
\end{figure}

\begin{figure}[!h]
  \centering
  \includegraphics[width=\linewidth]{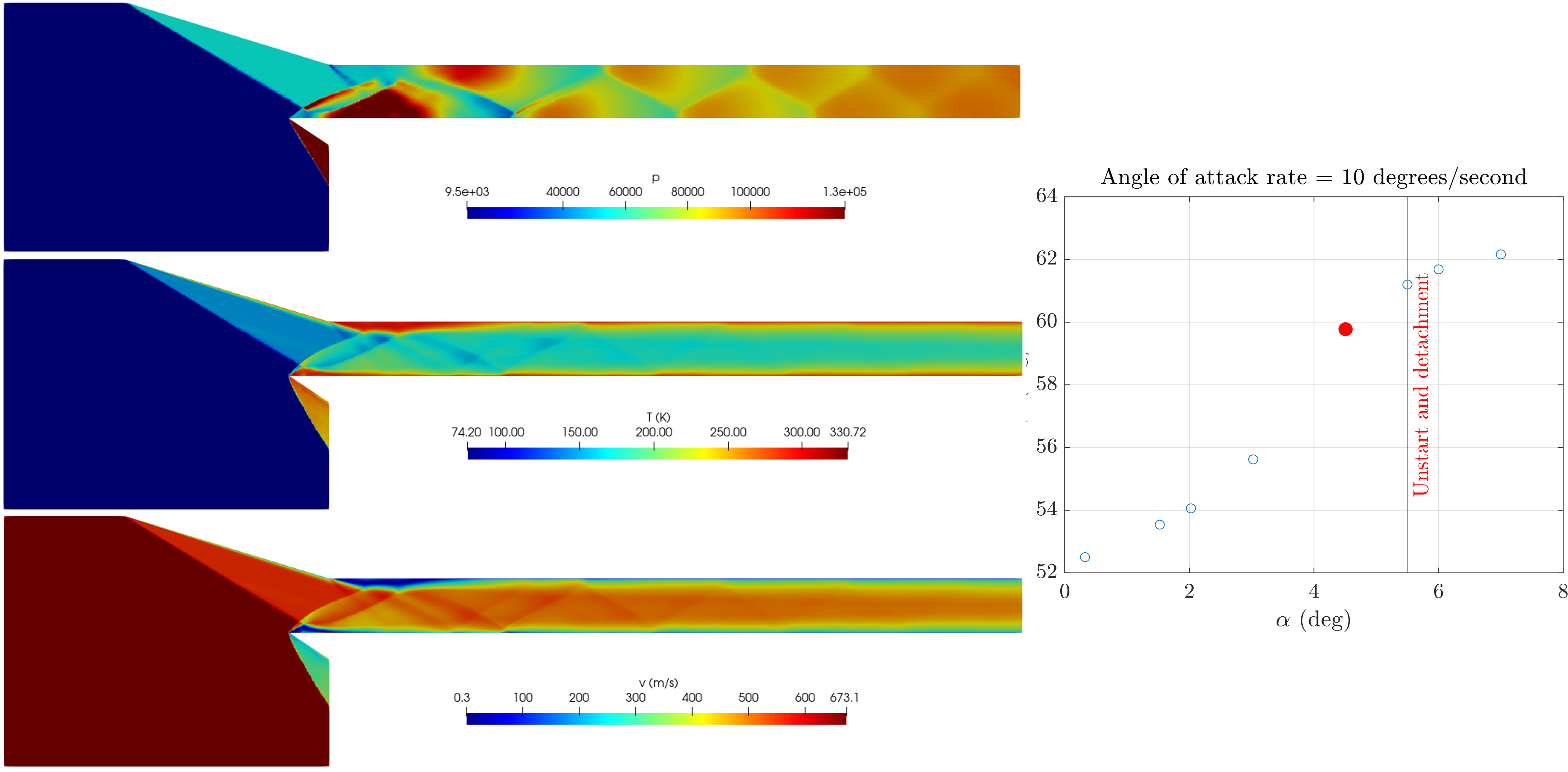}
  \caption{For the 10 deg/second angle of attack case, the pressure (top), temperature (middle), and velocity (bottom) contours at $\alpha = 4.5 \degree$ (red dot on the plot); relative oblique angle $\beta$ measured the same as in \autoref{fig:a10_61}, the shock generated from the point LLE, given on the right}
  \label{fig:a10_181}
\end{figure}

\begin{figure}[!h]
  \centering
  \includegraphics[width=\linewidth]{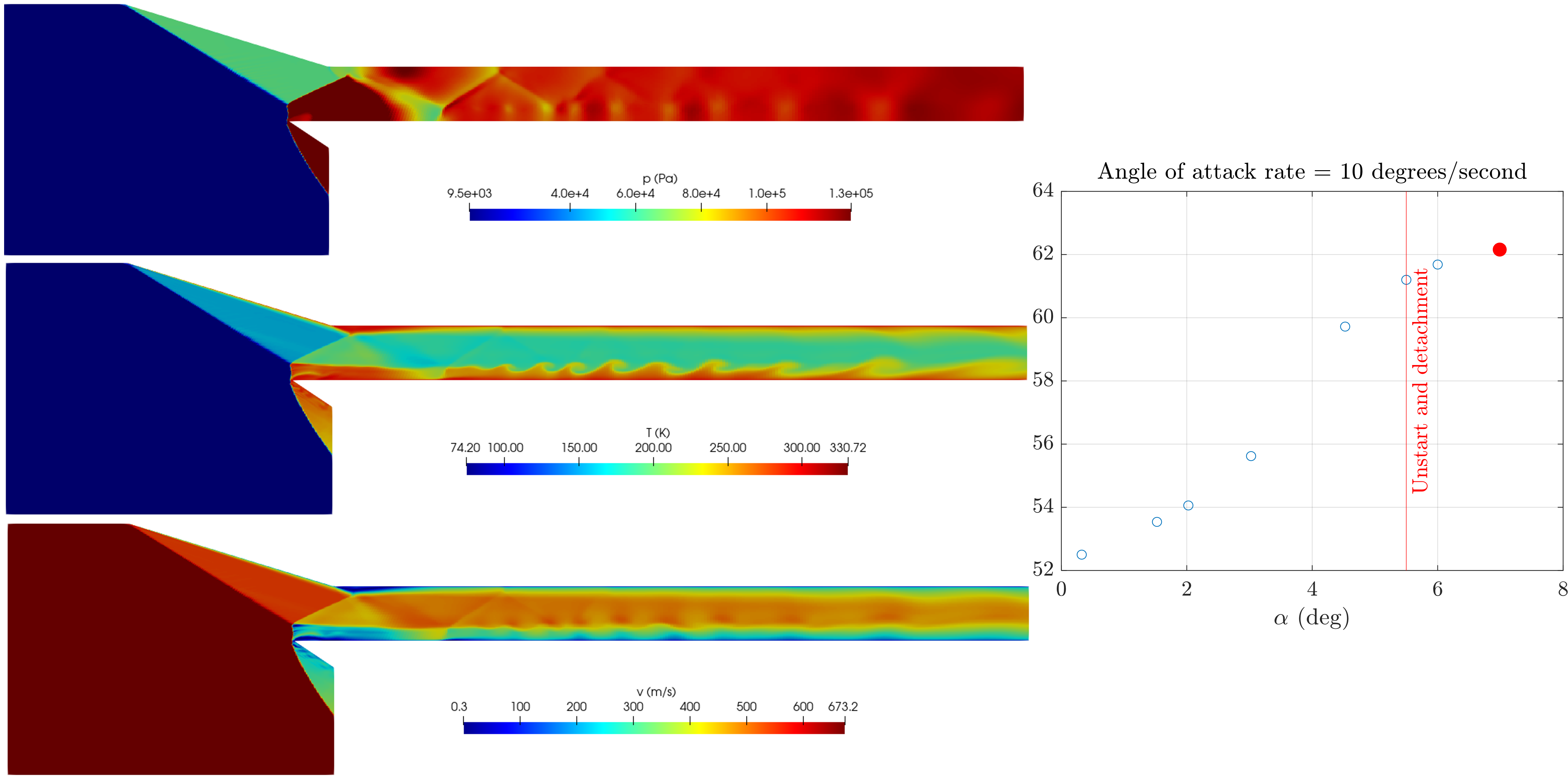}
  \caption{For the 10 deg/second angle of attack case, the pressure (top), temperature (middle), and velocity (bottom) contours at $\alpha = 7.0 \degree$ (red dot on the plot); relative oblique angle $\beta$ measured the same as in \autoref{fig:a10_61}, the shock generated from the point LLE, given on the right}
  \label{fig:a10_280}
\end{figure}

\begin{figure}[!h]
  \centering
  \includegraphics[width=\linewidth]{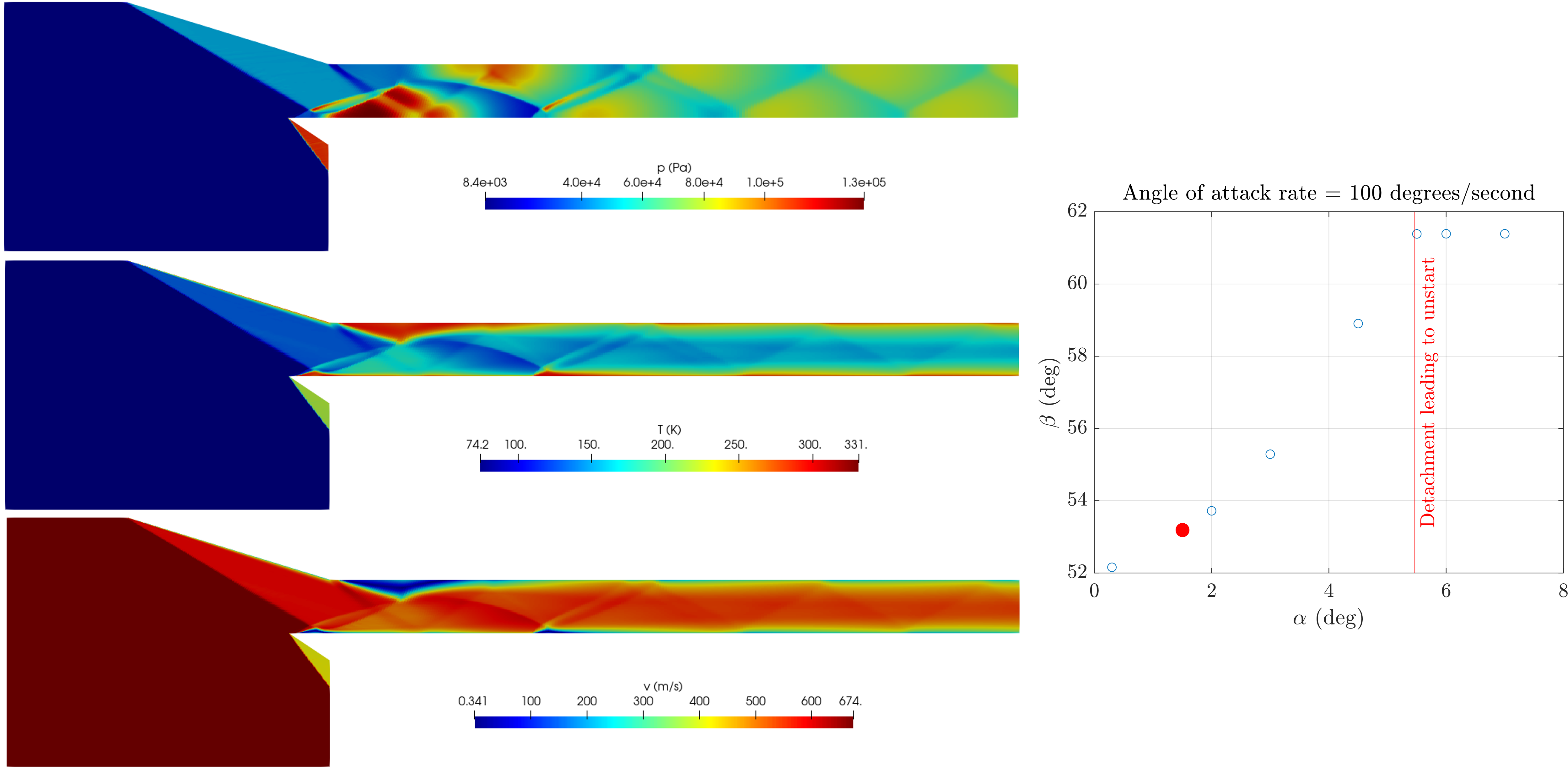}
  \caption{For the 100 deg/second angle of attack case, the pressure (top), temperature (middle), and velocity (bottom) contours at $\alpha = 1.5 \degree$ (red dot on the plot); relative oblique angle $\beta$ measured the same as in \autoref{fig:a10_61}, the shock generated from the point LLE, given on the right}
  \label{fig:a100_15}
\end{figure}

\begin{figure}[!h]
  \centering
  \includegraphics[width=\linewidth]{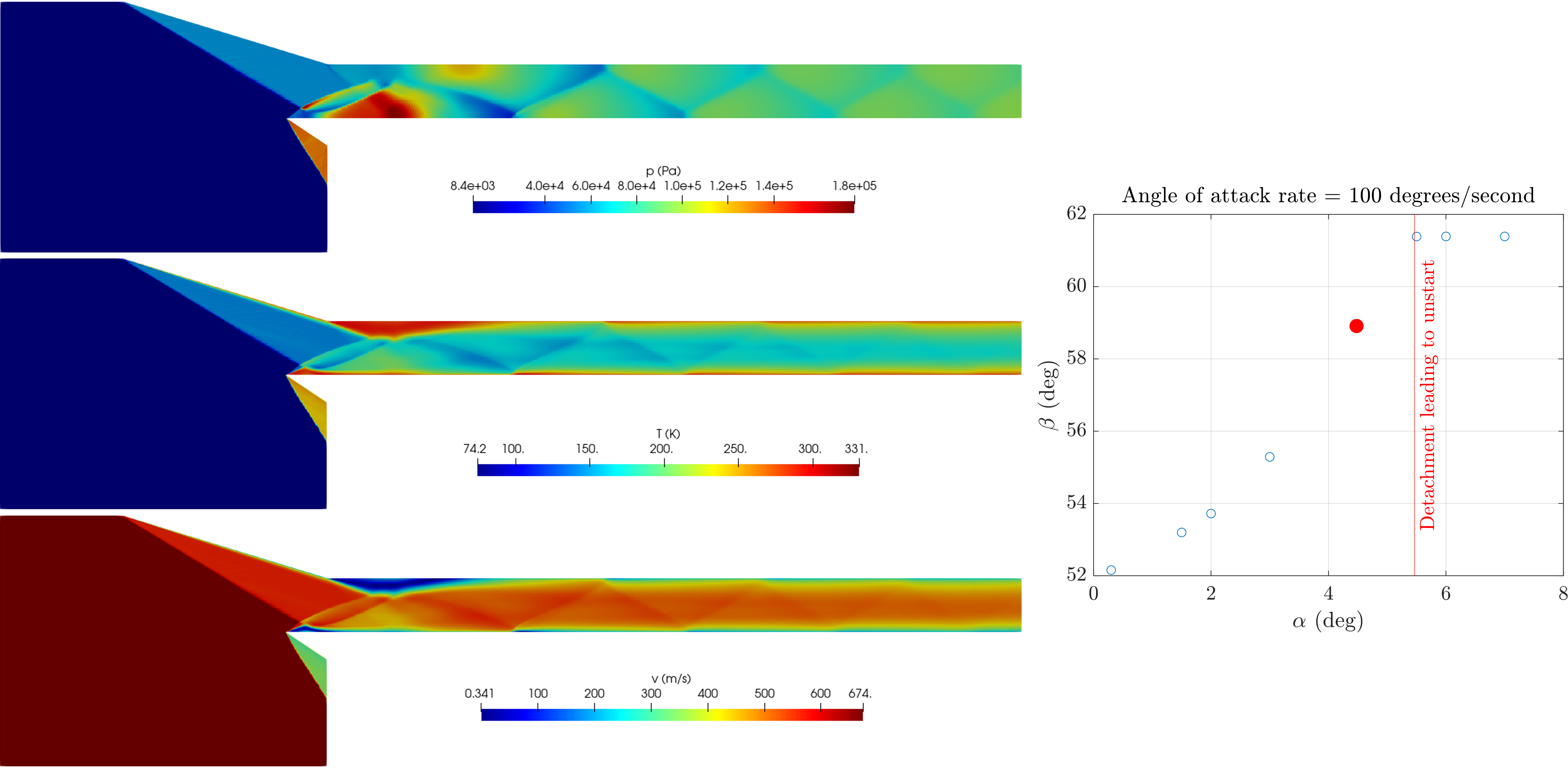}
  \caption{For the 100 deg/second angle of attack case, the pressure (top), temperature (middle), and velocity (bottom) contours at $\alpha = 4.5 \degree$ (red dot on the plot); relative oblique angle $\beta$ measured the same as in \autoref{fig:a10_61}, the shock generated from the point LLE, given on the right}
  \label{fig:a100_45}
\end{figure}

\begin{figure}[!h]
  \centering
  \includegraphics[width=\linewidth]{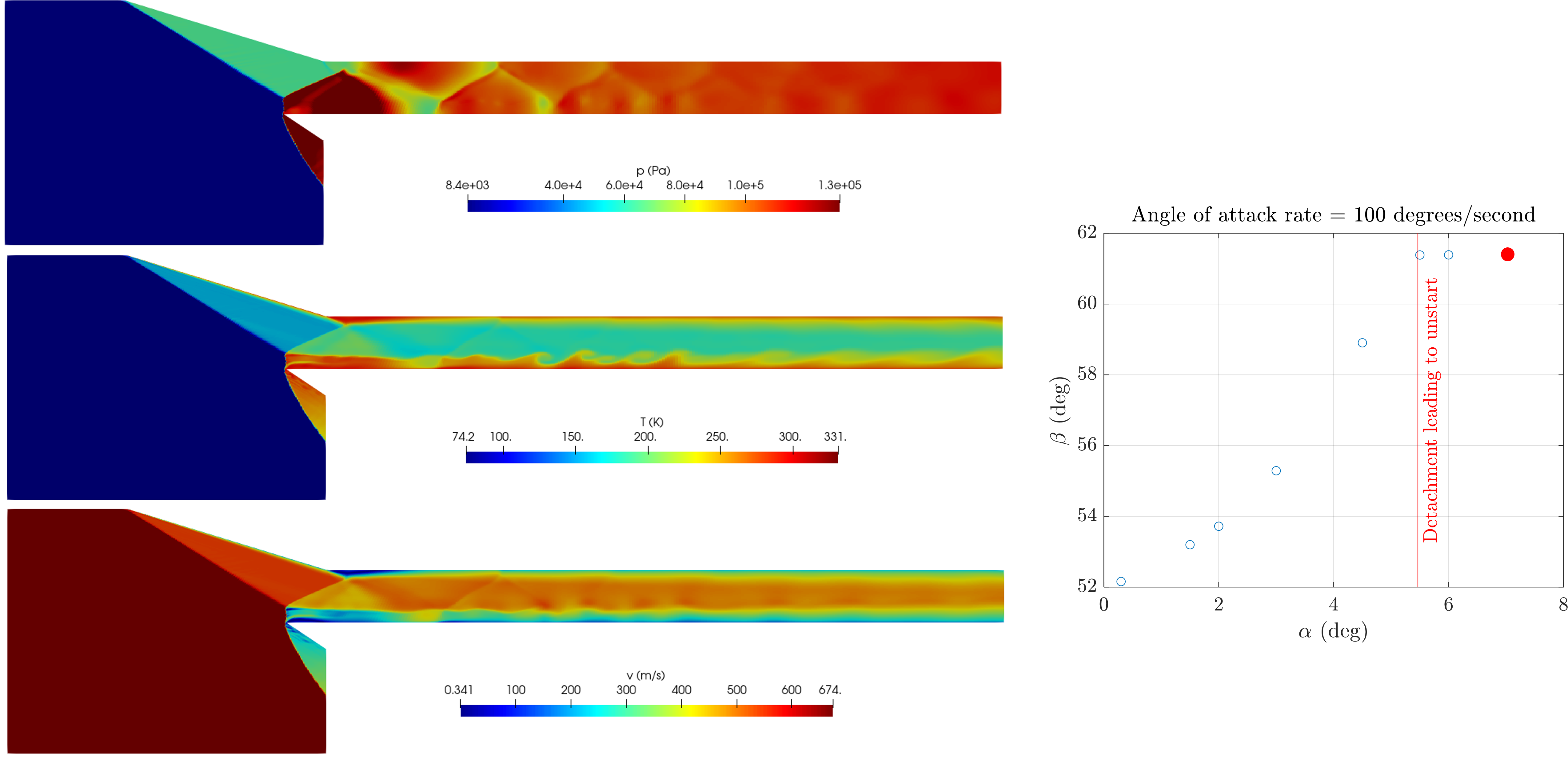}
  \caption{For the 100 deg/second angle of attack case, the pressure (top), temperature (middle), and velocity (bottom) contours at $\alpha = 7.0 \degree$ (red dot on the plot); relative oblique angle $\beta$ measured the same as in \autoref{fig:a10_61}, the shock generated from the point LLE, given on the right}
  \label{fig:a100_70}
\end{figure}

Iterations of pressure, temperature, and velocity of the flowfield solution are shown in figures \ref{fig:a10_61} -\ref{fig:a100_70}. The included image shows the progression towards unstart including the behavior noticed in the last sequence of images for each of the two rate cases. In the last of these images, unstart behavior is noticed including significantly high pressures, shock detachment, a breakdown of shock structures, and subsonic kelvin-helmholtz-type instabilities. It is critical to note here that the red vertical line represents the beginning of unstart in each case, but plots are shown at specific angles of attack to be able to compare one set to the other. \\

In visualizing the pressures over time for the 100 degree/s case, compression within the shock train is observed over time. The shock train in here is greatly effected by the recirculation region at the expansion into the inlet at point (UIE). The shape of the shock trains form something like an x-type shock train, however they are very close to being in sync. Not only that, but one of the interacting shock trains, the one initiated at point (LLE) entering the isolator, moves toward the inlet at a slightly faster rate, sometimes causing a violent shaking in the shock train, when the two trains overlap temporarily.\\

This phenomena is observed due to in part due to shock train interaction with differences in the leading edge geometries on the upper and lower surfaces, but also on the interaction of these shockwaves with the recirculation region. Any shocktrain that would enter this region is effectively dissipated, and instead, the shock waves are observed reflecting off of this region, sometimes causing further reflections based on the shape of the recirculation zone at that instant. By moving upstream at a higher rate, the lower leading edge shock wave moves past the lowest point of the recirculation zone, causing further separation between the two shock waves, where the most distinct x-type structure is observed. By the time the shock train from the upper leading edge catches up, detachment from the lower leading edge has begun, and the shock trains are effectively coalesced. In the 10 degree/s case, a similar phenomena is observed, but the system has a longer time to respond to the interaction of the lower leading edge shock train and the recirculation zone, causing a much higher pressure just below that zone. The rate differences betweeen the traveling speed of the leading shockwave in the system is computed as, 

\begin{equation} \label{eq2}
\begin{cases}

   \text{Case 1,} & v_{train} = -0.11 \text{ m/s} \\
   \text{Case 2,} & v_{train} = -1.08 \text{ m/s} 

    \end{cases}
\end{equation}\\

Comparing the final figures of the two cases, \autoref{fig:a10_280} and \autoref{fig:a100_70}, there are a few noticeable differences: (1) pressures are significantly higher in the 10 deg/s case than in that of the 100 deg/s case at the same aoa, (2) subsonic instabilities observed in the temperature and velocity plots are more developed and representative of a uniform Kelvin Helmholtz instability in the 10 deg/s case (3) temperature profiles are similar in magnitudes but not in trends (change significantly near the KH instabilities). These three observations give further credence to the theory that inlet/isolator unstart is not only dependent on angle of attack, but also on the rate of change of angle of attack. The pressures, velocities, and temperatures within the flowfield are unable to respond quickly enough to the changes in the 100 deg/s case, causing highly unsteady and underdeveloped flow features which if left unaccounted for in a rapid maneuver could lead to unstart much earlier than expected. \\


\clearpage

\subsubsection{10 deg/sec}

\subsubsection{100 deg/sec}

\clearpage

\section{Summary and Conclusions}

This study utilized the Eilmer state-of-the-art compressible Navier-Stokes equations solver to analyze unstart phenomena in a scramjet inlet due to angle of attack. The classical Spalart-Almaras RANS turbulence modeling approach was employed. Simulations were conducted for the NCSU planar inlet with a Mach 2.5 shock-on-lip design and a contraction ratio of 2.64. Subsequently, a series of simulations were performed at varying angles of attack and two distinct pitch rates (10 deg/s and 100 deg/s) to identify unstart onset. In both cases the timing of inlet unstart is observed to correlate well with the rapid surge in exit pressure as well as shock detachment
at the lower leading edge region. Lastly, exit pressures are significantly higher in the 10 deg/s case than in that of the 100 deg/s case at the same angle of attack. These observations suggest that unstart is not only dependent on angle of attack but also on AoA pitch rate. The findings offer valuable understanding of the unsteady behavior and flow evolution during hypersonic inlet unstart, with potential applications for unstart detection at high angles of attack.\\

Future work will investigate increasing the level of fidelity by extending to 3D and employing LES modeling techniques. Other areas of interest include conducting further validation efforts with experiments as well as exploring unstart mitigation strategies.


\section*{Acknowledgments}
This research was supported by DEVCOM Army Research Laboratory grants, W911NF2320026. Jeremy Redding is a PhD Fellow supported through the Army Educational Outreach Program (AEOP) program with ARL, cooperative agreement W9115R-15-2-0001. Luis Bravo and Muthuvel Murugan were supported by the 6.1 basic research program in propulsion sciences. The authors also gratefully acknowledge Dr. Jack Edwards for sharing his experience on modeling scramjet internal flow physics. The authors gratefully acknowledge the High-Performance Computing Modernization Program (HPCMP) resources and support provided by the Department of Defense Supercomputing Resource Center (DSRC) as part of the 2022 Frontier Project, Large-Scale Integrated Simulations of Transient Aerothermodynamics in Gas Turbine Engines. The views and conclusions contained in this document are those of the authors and should not be interpreted as representing the official policies or positions, either expressed or implied, of the DEVCOM Army Research Laboratory or the U.S. Government. The U.S. Government is authorized to reproduce and distribute reprints for Government purposes notwithstanding any copyright notation herein.

\bibliography{refs/sample.bib}

\end{document}